\documentstyle[seceq,supplement,epsf,wrapft,mbf]{ptptex}

\newcommand{\beq}{\begin{equation}}
\newcommand{\eeq}{\end{equation}}
\newcommand{\beqa}{\begin{eqnarray}}
\newcommand{\eeqa}{\end{eqnarray}}
\newcommand{\ba}{\begin{array}}
\newcommand{\ea}{\end{array}}

\pubinfo{~~~{\bf Preprint CAMTP/99-8}}  
\nofigureboxrule
\notypesetlogo  
\preprintnumber[3cm]{
CAMTP/99-8}

\begin{document}

\title{
High order WKB prediction of the energy splitting in the symmetric double 
well potential
}

\author{
Marko {\sc Robnik}\footnote{E-mail address: robnik@uni-mb.si},
Luca {\sc Salasnich}$^{*,}$\footnote{E-mail address: salasnich@mi.infm.it} and
Marko {\sc Vrani\v car}\footnote{E-mail address: mark.vranicar@uni-mb.si}

}

\inst{
CAMTP Center for Applied Mathematics and Theoretical Physics,
\\University of Maribor, Krekova 2, SI-2000 Maribor, Slovenia\\
$^*$Istituto Nazionale per la Fisica della Materia, 
 Unit\`a di Milano, \\
 Dipartimento di Fisica, Universit\`a di Milano, \\
 Via Celoria 16, I-20133 Milano, Italy
}
   

\abst{
The accuracy of the WKB approximation when predicting the 
energy splitting of bound states in a double well potential
is the main subject of this paper.
The splitting of almost degenerate energy levels below the top of the 
barrier results from the tunneling and is thus supposed to be exponentially
small.
By using the standard WKB quantization 
we deduce an analytical formula 
for the energy splitting, 
which is the usual Landau formula 
with additional quantum corrections. 
We also examine the accuracy  
of our and Landau formula numerically
for the case of
the symmetric double well quartic potential.
}

\maketitle

\section{Introduction}
Apart from the perturbative and variational methods the semiclassical or WKB 
approximation is the most widely used approximation in quantum mechanics 
to obtain the analytic expressions.
The semiclassics serves not only as an approximative tool but it also helps us
to picture and  to understand the global 
behaviour of eigenfunctions and energy spectra of quantum systems
which is particularly important in the context of quantum 
chaos\cite{casati,gutzwiller}.
Since the systematic study of the accuracy of semiclassical approximation 
is a difficult task, it has been 
attempted for simple systems, where in a few cases 
even exact solutions may be worked out\cite{bender,voros,rs1,rs2,ss}.

In this paper we analyze the energy splitting 
of doublets in a generic one--dimensional double well potential. 
The splitting due to the tunneling is supposed to be exponentially small and
is thus  lost a-priori  in the semiclassical approximation by expanding the 
quantum phase in the $\hbar$ power series.
By using the standard WKB technique  we deduce an analytical formula 
for the energy splitting which is the usual Landau\cite{landau}  formula 
with additional quantum corrections and can be formally written as
\beq\label{de1} 
\Delta E = A \exp{\Big[-{S\over \hbar }\Big]} \, ,
\eeq
where $S$ is the usual classical action 
inside the classically forbidden region (between the two 
turning points) 
and $A$ is called the tunneling amplitude, which can be written as an 
$\hbar$ power series. 
This formula is based on a linear approximation of the 
potential near the turning points and so is inapplicable in the case of the
square potentials\cite{rsv}. 
First 
we introduce the basic definitions, then 
we present the WKB splitting formula based on the higher order semiclassical 
WKB expansion and finally we study its  
validity for the quartic potential. 
We shall
demonstrate that our formula is indeed a significant improvement
of the Landau approximation\cite{landau}. Also, we shall
mention some potential applications in the field of molecular physics.

\section{Basic formalism}
Let us consider a particle in a one--dimensional 
symmetric double--well potential. 
The stationary Schr\"odinger equation of the system reads as
\beq\label{sch1} 
{\hat H} \psi (x) = \Big( -{\hbar^2\over 2m} {d^2 \over dx^2} + V(x) \Big) 
\psi (x) = E \psi (x) \, . 
\eeq
The Sturm--Liouville theorem (see, for example\cite{ch}) 
ensures that for one--dimensional systems 
there are no degeneracies in the spectrum. Let $\psi_1$ and $\psi_2$ be 
two exact and  almost degenerate 
eigenfunctions of the Schr\"odinger equation corresponding to the
energy eigenvalues $E_1$ and $E_2$ lying below the top of the 
barrier. 
We write the eigenfunctions $\psi_1$ and $\psi_2$ in terms 
of the right and left localized functions 
\beq\label{ef1} 
\psi_0(x) = {1\over \sqrt{2}}(\psi_1(x) + \psi_2(x) )
~~~~~~{\rm and }~~~~~~
\psi_0(-x) = {1\over \sqrt{2}}(\psi_1(x) - \psi_2(x) )\, .  
\eeq\label{de2} 
It is easy to show that $E_0 = <\psi_0(x) | {\hat H} |\psi_0(x) >=
 <\psi_0(-x) | {\hat H} |\psi_0(-x) >= {1\over 2} (E_1 + E_2)$ and 
\beq
\Delta E =E_2-E_1= {2\hbar^2 \over m} \psi_0(0)\psi{'}_0(0) \, ,
\eeq
(see\cite{landau,rsv}).
We shall remark that so far
this is an almost exact starting 
formula to calculate the energy splitting, since 
the error committed in the derivation is only exponentially 
small\cite{landau,rsv}. 
Please note that this quantity is always positive,
because the tail of the right localized eigenfunction
$\psi_0(x)$ at $x=0$ has the same sign for $\psi_0(0)$
and its derivative $\psi'_0(0)$.

\section{Semiclassical method}

The WKB splitting formula is obtained by inserting the WKB approximant for
the right localized function $\psi_0$ which is determined by 
the  WKB expansion of the Schr\"odinger equation\cite{mf}. 
A generic eigenfunction $\psi$ of the Schr\"odinger equation 
can always be written as
\beq\label{semef1} 
\psi (x) = \exp{ \big( {i\over \hbar} \sigma (x) \big) } \, ,
\eeq
where the phase $\sigma (x)$ is a complex function that satisfies 
the Riccati differential equation
\beq\label{ricati} 
\sigma{'}^2(x) + ({\hbar \over i}) \sigma{''}(x) = 2m(E - V(x)) \, .
\eeq
The WKB expansion for the phase is given by
\beq\label{semser} 
\sigma (x) = \sum_{k=0}^{\infty} ({\hbar \over i})^k \sigma_k(x) \, .
\eeq 
Substituting (\ref{semser}) into (\ref{ricati}) 
and comparing like powers of $\hbar$ gives 
the recursion relation ($n>0$)\cite{bender} 
\beq\label{rec} 
\sigma{'}_0^2=2m(E-V(x)) \, , \;\;\;\; 
\sum_{k=0}^{n} \sigma{'}_k\sigma{'}_{n-k}
+ \sigma{''}_{n-1}= 0 \, .
\eeq
The first five orders in the 
WKB expansion are given in the Appendix.

In particular, if we call $a$ and $b$ the two turning points 
(let $a<b$) corresponding to the energy $E$, 
the right localized wavefunction $\psi_0$ for the forbidden region $x<a$ 
that enters the splitting formula is given by
\beq\label{semef2} 
\psi_{0}(x) = 
\frac{C}{\sqrt{|\tilde{p}|}}\exp\left[\frac{1}{\hbar}\left(\int_a^x
|\tilde{p}|\,dx+\tilde{\sigma}_{even}+\tilde{\sigma}_{odd}\right)\right] \, ,
\eeq
where 
\beq\label{semef2a} 
\tilde{\sigma}_{even}=\sum_{k=1}^{\infty}\hbar^{2k}
\underline{\tilde{\sigma}_{2k}}
(|\tilde{p}(x)|);~~~
\tilde{\sigma}_{2k}(x)=
\underline{\tilde{\sigma}_{2k}}(|\tilde{p}(x)|)=\int_a^x\underline{\tilde{\sigma}'_{2k}}
(|\tilde{p}(\xi)|)d\,\xi \, ,
\eeq
$$
{\rm with}~~~
\underline{\tilde{\sigma}_{2k}}(-|\tilde{p}|)=-\underline{
\tilde{\sigma}_{2k}}(|\tilde{p}|)
\, ,
$$
and
\beq\label{semef2b}
\tilde{\sigma}_{odd}=\sum_{k=1}^{\infty}\hbar^{2k+1}\underline{\tilde{\sigma}_{2k+1}}
(|\tilde{p}(x)|);~~~
\tilde{\sigma}_{2k+1}(x)=
\underline{\tilde{\sigma}_{2k+1}}(|\tilde{p}(x)|)=
\int^x\underline{\tilde{\sigma}'_{2k+1}}(|\tilde{p}(\xi)|)d\,\xi
\eeq
$$
{\rm with}~~~
 \underline{\tilde{\sigma}'_{2k+1}}(-|\tilde{p}|)=
\underline{\tilde{\sigma}'_{2k+1}}
(|\tilde{p}|)\, ,
$$
where $\tilde p=\sqrt{2m(V(x)-E)}$ and if we define
$\sigma'_k(x)=\underline{\sigma'_k}(p(x))=f_k(p(x))$ then
$\tilde \sigma'_k(x)=\underline{\tilde\sigma'_k}(\tilde p(x))
=f_k(\tilde p(x))$, for  $f_k$ being the same function of an appropriate 
argument in both cases and $\underline{\tilde{\sigma}_{k}}
(|\tilde{p}(x)|)$ is considered as a family of functionals.
In evaluating the integrals $\sigma_k$ and $\tilde\sigma_k$ we cannot
integrate naively on the real axis, because such integrals
are divergent, but must take a partial derivative w.r.t.
the energy of certain fundamental complex contour integral\cite{bender}.
For potentials with only one minimum, thus having only two turning points, 
this has been proven by Robnik and Romanovski\cite{rr2}.

The constant $C$ is determined by the Kramers correspondence 
formula\cite{merzbacher} and
normalization condition for
the right localized function 
\beq\label{norm1}
1 = \int_{0}^{\infty} |\psi_0(x)|^2 dx \, ,
\eeq
which reads as 
\beq\label{norm2}
2 C^2 \int_a^b \; {1\over p} 
\exp{\big(2{\sigma_{odd}\over \hbar}\big)} dx = 1 \, , 
\eeq
that is approximately an integral of the semiclassical probability density
$|\psi_0(x)|^2$ over the classically allowed region $a<x<b$\cite{rsv}.

In this way the splitting formula, up to the 5th order, after taking into
account straightforward recursion relation (\ref{rec}) for $\sigma'_k$ or 
$\tilde\sigma'_k$ becomes
\beqa\label{de3}
\Delta E &  = & \frac{\hbar}{m}\left[\int_a^b\frac{1}{p}\,dx\right]^{-1}  
                \exp\left[-\frac{2}{\hbar}\int_0^a|\tilde{p}|\,dx\right] 
                \left\{\frac{}{}1+2\hbar\tilde{\sigma}_2(0)
                +2\hbar^2\left(\tilde{\sigma}_2^2(0)+
                \frac{\int_a^b\frac{\sigma_3}{p}\,dx}
                 {\int_a^b\frac{1}{p}\,dx}\right) \right. \nonumber \\ 
         & &      +2\hbar^3\left(\tilde{\sigma}_4(0)+
               \frac{2}{3}\tilde{\sigma}_2^3(0)+2\tilde{\sigma}_2(0)
               \frac{\int_a^b\frac{\sigma_3}{p}\,dx}
               {\int_a^b\frac{1}{p}\,dx}\right)     
          +2\hbar^4\left(\frac{1}{3}\tilde{\sigma}_2^4(0)+
                2\tilde{\sigma}_2(0)\tilde{\sigma}_4(0)\right) 
                \nonumber \\[2mm]
         &   &  +\left.2\hbar^4\left(2\tilde{\sigma}^2_2(0)
                \left[\frac{\int_a^b\frac{\sigma_3}{p}
               \,dx}{\int_a^b\frac{1}{p}\,dx}\right] 
              +2\left[\frac{\int_a^b\frac{\sigma_3}{p}\,dx}
            {\int_a^b\frac{1}{p}\,dx}\right]^2-
            \frac{\int_a^b\frac{\sigma^2_3+\sigma_5}{p}\,dx}
            {\int_a^b\frac{1}{p}\,dx}\right)\right\} 
\eeqa 
This formula is the usual Landau\cite{landau} formula for the energy splitting 
(1st order in $\hbar$ for the tunneling amplitude) 
with additional quantum corrections (up to the 5th order 
in $\hbar$ for the tunneling amplitude). We note that 
higher--order WKB corrections quickly increase in complexity\cite{rs1,rs2}
but, in principle, 
they can be calculated from the equation (\ref{rec}). 
It is important to stress that 
our splitting formula is good if the potential is sufficiently smooth 
so that the linear approximation and so the Kramers correspondence
relations are
valid near the turning points. 

The same splitting formula (\ref{de3}) can be derived using the
semiclassical scattering formalism and Kramer correspondence rules, 
for example as expounded
by Iyer and Will\cite{iw} and by  Will and Guinn\cite{wg}.
However, it is necessary
to go back to the very first step in their formalism to get
the result (\ref{de3}) of our problem. We have done this and confirmed, as 
mentioned above, that the result is the same.

\section{Quartic potential}

As one of our test models we consider the symmetric double well
quartic potential, given by 
\beq\label{qp1}
V(x)=-2Bx^2+Ax^4 \; ,
\eeq
where the parameters $A$ and $B$ are related to the potential 
barrier $V_0$ and to the position of the minimum $x_0$ by 
\beq\label{qp2}
V_0=\frac{B^2}{A},~~~x_0=\sqrt{\frac{B}{A}} \, .
\eeq
By using the following reduced variables
\beq\label{qp3}
\bar{x}=\frac{x}{x_0} \; ,~~~\bar E=\frac{E}{V_0}\, , 
~~~\hbar_{eff}=\frac{\hbar}{x_0\sqrt{mV_0}} \, ,
\eeq
the quantum Hamiltonian operator of the system can be written as
\beq\label{q1}
{\hat H}= -{\hbar_{eff}^2\over 2}{\partial^2 \over \partial \bar{x}^2} 
-2\bar{x}^2+\bar{x}^4 \, . 
\eeq

The exact spectrum of operator (\ref{q1}) is calculated by 
numerical diagonalization of the Hamiltonian matrix, written in the 
basis of harmonic oscillator in 
QUADRUPLE PRECISION number format (32 digits)\cite{rsv}.

The quantities $\tilde{\sigma}_0(0)$, $\tilde{\sigma}_2(0)$, $\tilde{\sigma}_4(0)$,
$\int_a^b\frac{1}{p}\,dx$, $\int_a^b\frac{\sigma_3}{p}\,dx$ and
$\int_a^b\frac{\sigma^2_3+\sigma_5}{p}\,dx$ that enters the semiclassical 
splitting formula can all be expressed in terms of complete elliptic integrals of 
the first and the second kind\cite{rsv}.

Please note that all the integrands of these quantities are strongly 
divergent at the turning points, but all the expressions 
in (\ref{de3}) can be made convergent and finite by taking partial derivatives 
with respect to $E$ of certain finite expressions\cite{bender,rs1,rs2,rsv}.

\begin{wrapfigure}{r}{7cm}
\epsfxsize=7.0cm
\epsfbox{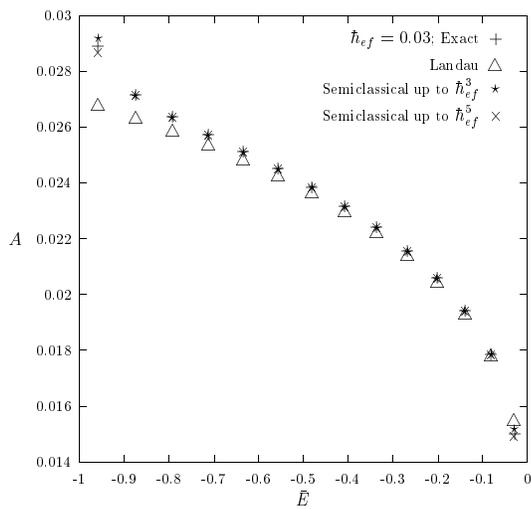}
\caption{Tunneling amplitude $A$ {\it vs} mean energy $\bar{E}$ of 
Quartic potential with $\hbar_{eff}=0.03$.}
\end{wrapfigure}
In figure 1 we show the tunneling amplitude $A$ as a function 
of the mean energy $\bar{E}$ of all almost degenerate pairs of
quartic potential with $\hbar_{eff}=0.03$. 
We compare the exact results 
with the semiclassical ones at 1st (Landau), 3rd and 5th 
order in $\hbar$. 
We can observe that, as expected, the agreement is better at the energies
for which the well known semiclassical criterion is fulfilled. The 
criterion demands the absolute value of the de Broglie wavelength to be small
compared to the typical scale of the potential. This is not the case 
at energies near the top and at the bottom of the potential barrier but
the accuracy increases when going with the energy away from those two values.
In figure 2 (3) we plot the tunneling amplitude $A$ as a 
function $\hbar_{eff}$ for the first (fourth) pair of almost degenerate 
consecutive energy levels. As shown also in table 1, 
the semiclassical results approach the exact ones by increasing 
the perturbative order in $\hbar$. Note that at the 5th order in $\hbar$ 
the agreement with exact result is up to the 8th digit. 

\begin{figure}[h]
\parbox{7cm}{
\figurebox{7cm}{0cm}
\epsfxsize=7cm
\centerline{\epsfbox{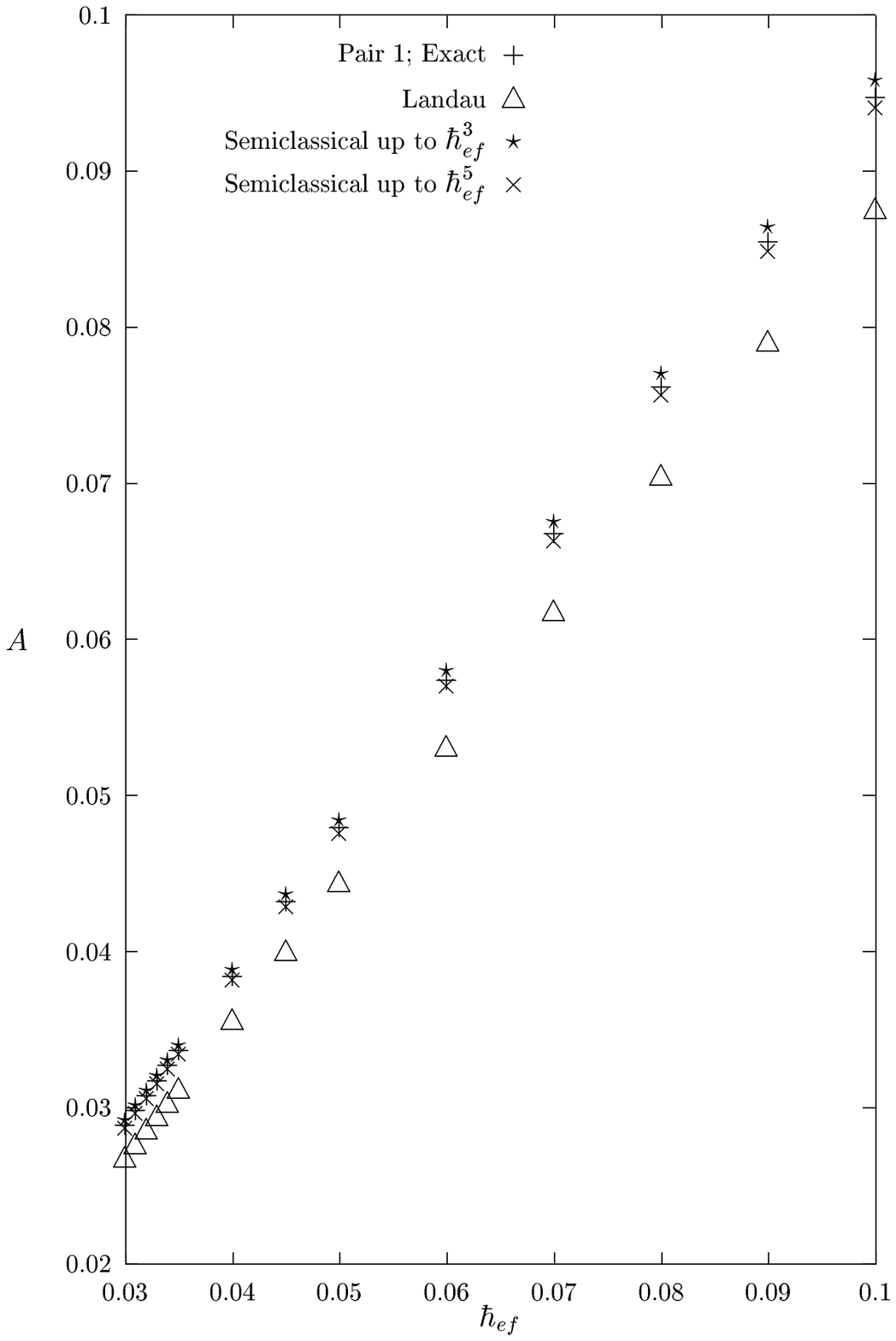}}}
\hspace{0mm} 
\parbox{7cm}{
\figurebox{7cm}{0cm}
\epsfxsize=7cm
\centerline{\epsfbox{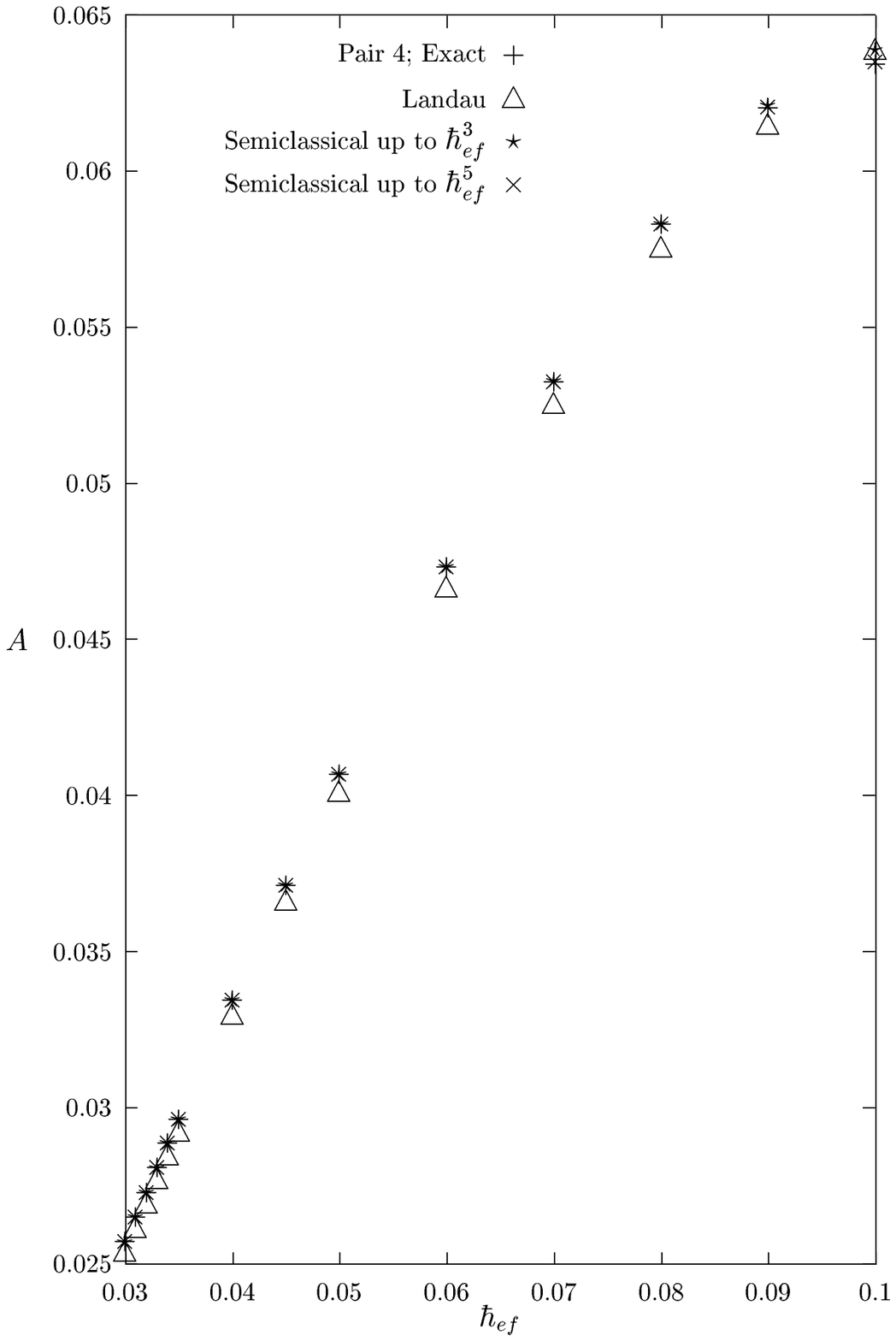}}}
\caption{Tunneling amplitude $A$ {\it vs} $\hbar_{eff}$ for 
the first pair of almost degenerate consecutive energy levels (left). The same plot
for the forth pair of almost degenerate consecutive energy levels (right). 
Quartic potential. }
\label{fig3}
\vspace{-2mm}
\end{figure}

\begin{table}[]
\caption{Energy splitting $\Delta \bar{E}$ for all almost degenerate 
pairs at $\hbar_{eff}=0.03$. $\bar{E}_s$ is the exact mean energy, 
$\Delta\bar{E}_{Landau}$, $\Delta\bar{E}_{\hbar^3}$ and 
$\Delta\bar{E}_{\hbar^5} $ are the semiclassical results at 
1st (Landau), 3rd and 5th order, respectively. 
In the last column of the upper table: exact energy splitting 
in units of mean level spacing.}
\begin{center}
\footnotesize
\begin{tabular}{l|r|r|r} \hline\hline
 & & & \\[-3mm]
Pair number & $\bar{E}_s$ & $\Delta \bar{E}_{Exact}$ & 
$\Delta \bar{E}_{n}/(\overline{\bar{E}_s^{n+1}-\bar{E}_s^{n})}$ \\ \hline
1  & -0.9578013510838623 & 9.232(2)E-28            &  1.291(6)E-26         \\
2  & -0.8743363182686136 & 6.533698E-25         &  9.140541E-24        \\
3  & -0.7923227236203907 & 2.197647599E-22      &  3.074474586E-21  \\
4  & -0.7118393444104010 & 4.66709587543E-20     & 6.52919406381E-19   \\
5  & -0.6329771419061873 & 7.0097983270055E-18  &  9.8065981172742E-17   \\
6  & -0.5558426330893027 & 7.901944841994295E-16  &  1.10546971247045E-14  \\
7  & -0.4805627052945287 & 6.920784581323967E-14  &  9.68206928062435E-13   \\
8  & -0.4072917698084478 & 4.806630946488390E-12  &  6.72440144371474E-11   \\
9  & -0.3362229365299848 & 2.675674688376366E-10  &  3.74322699989559E-09    \\
10 & -0.2676066496781732 & 1.196767019706533E-08  &  1.67425832453064E-07   \\
11 & -0.2017846855780184 & 4.273586943459416E-07  &  5.97868122857108E-06    \\
12 & -0.1392610433553887 & 1.195177462312803E-05  &  1.67203455862220E-04    \\
13 & -8.088844930248588E-02 & 2.500127371158503E-04  &  3.49763905139656E-03 \\
14 & -2.855578436131249E-02 & 3.352190557610379E-03  &  4.68966210972998E-02   \\

\hline
\end{tabular}
\end{center}
\vspace{3mm}
\begin{center}
\footnotesize
\begin{tabular}{r|r|r} \hline\hline
$\Delta \bar{E}_{Landau}(\bar{E}_s)$ & $\Delta \bar{E}_{\hbar^3}(\bar{E}_s)$ & 
$\Delta \bar{E}_{\hbar^5}(\bar{E}_s)$ \\ \hline

8.56386072023299E-28       & 9.{\bf 3}3424576483334E-28 & 9.1{\bf 6}991849898935E-28 \\
6.{\bf 3}4511775769322E-25 & 6.53{\bf 7}96204696428E-25 & 6.533{\bf 3}2080224002E-25 \\
2.1{\bf 5}758453279967E-22 & 2.197{\bf 9}8578660872E-22 & 2.1976{\bf 3}766003663E-22 \\
4.6{\bf 0}423712282166E-20 & 4.667{\bf 3}7049838592E-20 & 4.66709{\bf 2}38092202E-20 \\
6.9{\bf 3}436277656352E-18 & 7.010{\bf 0}0176474235E-18 & 7.00979{\bf 7}08264343E-18 \\
7.8{\bf 3}077639810467E-16 & 7.902{\bf 0}7840692513E-16 & 7.901944{\bf 4}5833250E-16 \\
6.8{\bf 6}703907509728E-14 & 6.9208{\bf 6}233123417E-14 & 6.9207845{\bf 1}430764E-14 \\
4.7{\bf 7}381431395796E-12 & 4.8066{\bf 7}193943203E-12 & 4.8066309{\bf 8}878842E-12 \\
2.6{\bf 5}943820818162E-10 & 2.6756{\bf 9}528644699E-10 & 2.6756747{\bf 7}543535E-10 \\
1.19{\bf 0}31027734624E-08 & 1.1967{\bf 7}788146677E-08 & 1.196767{\bf 1}4995564E-08 \\
4.2{\bf 5}353592752249E-07 & 4.2736{\bf 5}612082292E-07 & 4.27358{\bf 9}20983707E-07 \\
1.19{\bf 0}79278916304E-05 & 1.1952{\bf 4}083756918E-05 & 1.19518{\bf 3}71126262E-05 \\
2.49{\bf 8}35207740361E-04 & 2.50{\bf 1}42635772383E-04 & 2.500{\bf 5}8564705645E-04 \\
3.{\bf 4}6270869995442E-03 & 3.3{\bf 9}489990444765E-03 & 3.3{\bf 3}278741702504E-03 \\

\hline
\end{tabular}
\end{center}
\vspace{-2mm}
\end{table}

\section{Conclusions}

In this work we have taken the first step towards a systematic
improvement of the Landau formula\cite{landau},
which is the semiclassical leading order energy level splitting
formula for pairs of almost degenerate levels in double well
potentials. We have developed the algorithm for the semiclassical
$\hbar$ expansion series to all orders for the tunneling
amplitude $A$ (of equation (\ref{de1})), and thus
calculated explicitly the quantum corrections
up to the 5th order. 
We have compared the semiclassical 
predictions with the exact results obtained numerically, 
for the  case of the quartic double well 
potential. Our approach is based on the usual WKB expansion
in one--dimensional potentials and so the 
calculation of higher corrections can in principle be continued by 
the same method, although the structure of higher terms
increases in complexity very quickly. We have also checked 
what happens in cases where the assumption implicit in the
Landau formula (namely the linearity of the potential around
the turning points) is not satisfied as in the case of double square well 
potential: We get a different
result even in the leading semiclassical order\cite{rsv}. 
We should stress that the Landau formula\cite{landau} is indeed quite good 
approximation since it always yields the correct order 
of magnitude (the exponential tunneling factor is always correct) and 
even the tunneling amplitude is correct within the $5$--$50$ \% . 

It is our goal to work out a more direct WKB approach 
to the solution of the multi--minima problem, 
by the contour integration technique, 
based on requiring the single valuedness of the eigenfunction,
as has been done by Robnik and Salasnich\cite{rs1,rs2} in the case of a single 
well potential. This is our future project.

Finally we should mention important applications e.g. in the
domain of molecular physics\cite{landau,herzberg,cdl}.  
For example, the $NH_3$ molecule can be 
described by a quasi one-dimensional potential $V(x)$ as a function of
the perpendicular distance  $x$ of the nitrogen  $N$ atom from the $H_3$ plane,
and as such it has the double well form, with the top of the
potential barrier at $x=0$. In order to calculate the energy
level splitting of the doublets of vibrational modes one
needs exactly our theory. Another example is the torsional
motion of $C_2H_4$ molecule, where again we encounter an
effectively one-dimensional double well potential. In case
of $C_2H_6$ molecule, we find three potential wells where
the tunneling effects again determine the splittings of  energy
triplet levels and a generalization of our theory would give an 
improved estimate of the splittings.

\section*{Acknowledgments}
This work was supported by the Ministry of Science and
Technology of the Republic of Slovenia and by the Rector's Fund of the 
University of Maribor. L.S. was partially supported 
by the Italian MURST (Prof. G. Cattapan).

\section*{Appendix}
The first 5 orders in the WKB expansion are:
\beqa\label{sigmas}
\sigma'_0 & = & p\, , \nonumber \\[3mm]
\sigma'_1 & = & -{p{'}\over 2p} \, , \nonumber \\[3mm]
\sigma'_2 & = & {p{''}\over 4 p^2} -{3\over 8}{p{'}^2\over p^3} \, , 
                \nonumber\\[3mm]
\sigma'_3 & = & \frac{p'''}{8p^3}+\frac{3}{4}\frac{p'p''}{p^4}-
             \frac{3}{4}\frac{p^{'3}}{p^5} \, , \nonumber \\[3mm]
\sigma'_4 & = & \frac{1}{16}\left(
             \frac{p''''}{p^4}-10\frac{p'''p'}{p^5}
             -\frac{13}{2}\frac{p^{''2}}{p^5}
             +\frac{99}{2}\frac{p''p^{'2}}{p^6}
             -\frac{297}{8}\frac{p^{'4}}{p^7}\right) \, , \\[3mm]
\sigma'_5 & = & \frac{1}{32} \left(
             -\frac{p'''''}{p^5}+15\frac{p''''p'}{p^6}
             +24\frac{p'''p''}{p^6} 
           -111\frac{p'''p^{'2}}{p^7}
              -144\frac{p^{''2}p'}{p^7}
           +510 \frac{p''p^{'3}}{p^8}
            -306\frac{p^{'5}}{p^9}\right) \nonumber \, .
\eeqa
The derivatives $\tilde\sigma'_k$-s can be obtained by replacing $p(x)$ by $\tilde p(x)$ 
in the expressions above.

\end{document}